\RequirePackage{lineno} 
\documentclass[preprint,aps]{revtex4}
\usepackage{amsmath}
\usepackage{graphicx}% Include figure files
\usepackage{dcolumn}% Align table columns on decimal point
\usepackage{bm}% bold math
\ruledtabular
\begin{document}
\title{Temperature induced transition from p-n to n-n electronic behavior in Ni$_{0.07}$Zn$_{0.93}$O/Mg$_{0.21}$Zn$_{0.79}$O heterojunction}
\author{Tanveer A. Dar$^{a}$}
\email{tanveerphysics@gmail.com}%agrawal.arpana01@gmail.com}
\author{Arpana Agrawal$^{b}$}\author{Ram J. Choudhary$^{c}$}\author{Pranay K. Sen$^{a}$}\author{Pankaj Misra$^{d}$}\author {Pratima Sen$^{b}$}
\affiliation{$^{a}$Department of Applied Physics and Optoelectronics, Shri. G. S. Institute of Technology and Science, Indore-452003, India}
\affiliation{$^{b}$School  of Physics, Devi Ahilya University, Takshashila Campus, Indore - 452001, India}
\affiliation{$^{c}$UGC-DAE Consortium for Scientific Research, Khandwa Road, Indore 452001, India}
\affiliation{$^{d}$Laser Materials Processing Division, Raja Ramanna Center for Advanced Technology, Indore-452013, India}
   
\begin{abstract}
 The transport characteristics across the pulsed laser deposited Ni$_{0.07}$Zn$_{0.93}$O/Mg$_{0.21}$Zn$_{0.79}$O heterojunction exhibits p-n type semiconducting properties at 10\,K while at 100\,K, its characteristics become similar to that of an n-n junction. The reason for the same is attributed to the role of larger electronegativity of Ni as compared to Mg at 10\,K and ionization of impurity states at 100\,K. The above behavior is confirmed by performing the Hall measurements.
\end{abstract}
%\linenumbers
\maketitle 

Non-ohmic transport characteristics across heterojunction diode are decisive ingredient for most of the contemporary technological applications.Accordingly, a wide range of p-type and n-type materials have been studied in combination. In recent years, much emphasis is given to the oxide based heterostructures displaying the rectifying behavior whose properties could be tuned by various parameters e.g. doping concentration, thickness, magnetic field, temperature etc. However, one of the major challenges that these heterostructures undergo is the mismatch of lattice parameters or crystalline structure of the chosen n/p-type layer material at the interface. This also leads to an abrupt change in electronic properties at the interface of such heterostructures due to which transport of the charge carriers are affected. In this context, efforts are being made to prepare p-n junction diode of the semiconductors of the same material by choosing appropriate dopants.

 Over last few decades, ZnO has emerged as a key oxide material with wide range of fascinating properties like direct wide band gap of 3.37\,eV and a large exciton binding energy of 60\,meV at room temperature \cite{Ozgur,Look}. n-type ZnO is readily available because of wide range of dopants available as well as inherent nature of ZnO itself which has a tendency to form n-type defect states such as oxygen vacancy or zinc interstitial \cite{Janoti}. Therefore, search is on to find a suitable dopant in ZnO, which can lead to p-type doping. However, so far the success is limited, and hence, the exploration for rectifying behavior is further extended to hybrid structures of n-type ZnO only, though with different carrier concentration in individual layer, such as n-n heterostructure \cite{Yoon}. The concentration gradient in such heterostructure at the interface will yield charge transfer across the junction and a potential barrier will develop depending on the dopant concentration profile and different band lineups. It should be noted here that the charge type responsible for the transport across the interface will not only depend on the nature of dopant and its concentration but also on the band alignment at the interface. Since, both energy bands and electron distribution are temperature sensitive; one can expect variation in the transport characteristics with respect to temperature \cite{A,B}. The temperature dependent transport studies of Zn$_{0.9}$Mn$_{0.1}$O/ZnO heterojunction showed that two-dimensional electron gas is successfully formed at the Zn$_{0.9}$Mn$_{0.1}$O/ZnO interface \cite{Edahiro}. Such systems are also useful for making temperature sensors \cite{C,Saiket}. 
 
 Recognizing the immense significance of the technological application of transport characteristics across a heterojunction under different environmental conditions, in the present letter, we have explored the temperature dependence of the transport characteristics across the Ni$_{0.07}$Zn$_{0.93}$O/Mg$_{0.21}$Zn$_{0.79}$O heterojunction.
%Non-ohmic transport characteristics across heterojunction diode occurs due to different band lineups or to the inequilibrium charge distribution across the interface. Since, both energy bands and electron distribution are temperature sensitive, one can expect variation in the transport characteristics with respect to temperature \cite{A,B}. Such a change is useful for making temperature sensors \cite{C,Saiket} but it limits the usability of the devices to a certain temperature regime \cite{GP}. However, the knowledge of temperature dependence of transport characteristics is important for making sophisticated electronic/optoelectronic devices. 

It is known that, Mg and Ni doped ZnO exhibit n-type conductivity at room temperature \cite{JAP,Shubhra} and their heterojunction can result in n-n junction. Doping ZnO with Mg causes an increase in the band gap of ZnO while Ni doping causes reduction of the band gap \cite{JAP, CAP}. So combining both, that is, Mg doped ZnO and Ni doped ZnO will result in enhancing the band offset at the interface. %Accordingly, we have calculated the band offset parameters in both the cases \cite{CAP,APL}. Also, from the transport characteristics across Mg$_{0.21}$Zn$_{0.79}$O/ZnO heterojunction, we observed anomalous resistance in resistance vs voltage (R-V) characteristics which were explained on the basis of occurrence of band offset at the heterojunction \cite{APL}.

 %The stimulus for the present work stems from the results reported in our earlier papers \cite{CAP,APL} suggesting that Ni$_{0.07}$Zn$_{0.93}$O/Mg$_{0.21}$Zn$_{0.79}$O interface would exhibit relatively a \textbf{larger band offset} and might result in the possibility of non-ohmic carrier transport across the heterojunction. 
%\section{Experiment}

For the present investigations we have performed the carrier transport characteristics across pulsed laser deposited (PLD) Ni$_{0.07}$Zn$_{0.93}$O/ Mg$_{0.21}$Zn$_{0.79}$O heterojunction. We ensured the same deposition conditions for depositing the single layer films of Ni$_{0.07}$Zn$_{0.93}$O and Mg$_{0.21}$Zn$_{0.79}$O as well as the heterojunction comprising of Ni$_{0.07}$Zn$_{0.93}$O/Mg$_{0.21}$Zn$_{0.79}$O bilayer film. 

The films were grown on quartz substrate by pulse laser deposition (PLD) technique using KrF excimer laser (330\,mJ). The pulse duration of the laser was 20\,ns with a repetition rate of 10\,Hz and the energy density was 2\,J/cm$^{2}$. Deposition was carried out in a vacuum chamber pumped down to a base pressure of 4$\times$10$^{-6}$\,mbarr and oxygen gas was flown into the chamber at 1$\times$10$^{-4}$\,mbarr. The target-to-substrate distance was maintained at 4\,cm with optimized substrate temperature of 400$^{o}$\,C. Ni$_{0.07}$Zn$_{0.93}$O layer (5\,nm) was developed on initially grown Mg$_{0.21}$Zn$_{0.79}$O film (200\,nm) by partially masking it. 

The grown samples were subsequently characterized by X-ray diffraction (XRD) for structural analysis. The current-voltage characteristics (I-V) across the junction were examined at 10\,K and 100\,K using standard linear four probe technique. For the investigation of carrier density and type of majority charge carriers at different temperature, we have carried out Hall measurements using Quantum design Physical Property Measurement System (Model 7100). Hall measurements were performed on the single layer Ni$_{0.07}$Zn$_{0.93}$O films at 10\,K and 100\,K. The magnetic field was applied perpendicular to the film plane and the current was kept fixed at 0.051\,mA. 

Figure 1 shows the XRD patterns of the Ni$_{0.07}$Zn$_{0.93}$O, Mg$_{0.21}$Zn$_{0.79}$O single layer films and Ni$_{0.07}$Zn$_{0.93}$O/Mg$_{0.21}$Zn$_{0.79}$O heterojunction. Observation of only (002) and (004) reflection peaks in all the samples suggests the growth of c-axis oriented films. Apart from these two peaks, no other peak corresponding to either Mg, Ni or their oxides is detected. Consequently, we infer that there is no impurity phase within the XRD detection limits and the grown samples are single phase retaining the wurtzite structure of ZnO.

\begin{figure} %XRD
\centering
\includegraphics[width = 0.55\columnwidth]{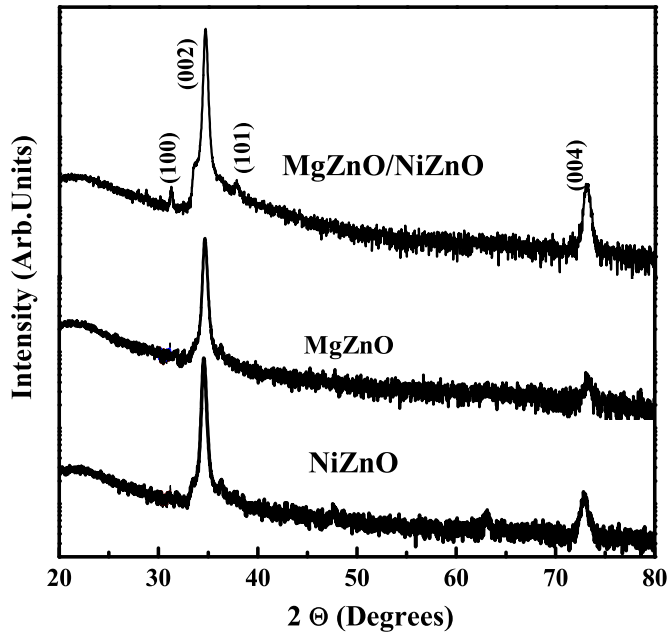}%
\caption{X-ray diffraction patterns of the grown samples.}% 
\label{XRD}%
\end{figure}
To get the insight about the charge transport across the Ni$_{0.07}$Zn$_{0.93}$O/Mg$_{0.21}$Zn$_{0.79}$O heterojunction, we have performed the I-V characteristics as shown in fig. 2. Interestingly, we observed that the heterostructure exhibits a typical ohmic behavior at 100\,K and a non-ohmic behavior at 10\,K. To investigate the origin of the existing non-ohmic behavior of the sample at 10\,K, we also measured the I-V characteristics of single layer Ni$_{0.07}$Zn$_{0.93}$O and Mg$_{0.21}$Zn$_{0.79}$O films near zero biasing regime, shown in fig 2(b) and 2(c). It has been observed that both the samples Ni$_{0.07}$Zn$_{0.93}$O and Mg$_{0.21}$Zn$_{0.79}$O show ohmic behavior at both the temperatures. Accordingly, we can conclude that the non-linearity at 10K arises only because of the junction properties of the grown sample. 

 %In order to investigate the charge transport across the Ni$_{0.07}$Zn$_{0.93}$O/Mg$_{0.21}$Zn$_{0.79}$O junction, we have measured its current voltage (I-V) characteristics at 10K and 100K. At higher temperatures, the junction does not show any significant nonlinearity in the I-V curve characteristics. The obtained results are shown in fig. 6.2 from which it is evident that the sample exhibits a typical ohmic behavior at 100K and a non-ohmic behavior is observed at 10K. To investigate the origin of the existing non-linear (non-ohmic) behavior of the sample at 10K, we also measured the I-V characteristics of single layer Ni$_{0.07}$Zn$_{0.93}$O and Mg$_{0.21}$Zn$_{0.79}$O films near zero biasing regime, shown in fig 6.2(b) and 6.2(c). It has been observed that both the samples Ni$_{0.07}$Zn$_{0.93}$O and Mg$_{0.21}$Zn$_{0.79}$O show linear (ohmic) behavior at both the temperatures. Accordingly, we can conclude that the non-linearity at 10K arises only because of the junction properties of the grown sample. 
\begin{figure} %Mgcore
\centering
\includegraphics[width = 0.55\columnwidth]{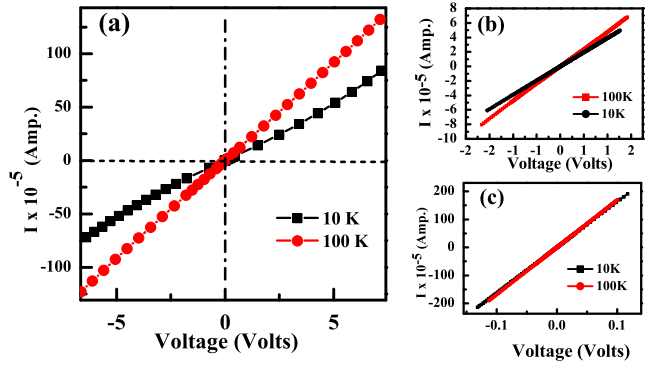}
\caption{Current-voltage relationship of (a) Ni$_{0.07}$Zn$_{0.93}$O/Mg$_{0.21}$Zn$_{0.79}$O heterojunction (b) Mg$_{0.21}$Zn$_{0.79}$O and (c) Ni$_{0.07}$Zn$_{0.93}$O at 10\,K and 100\,K.}% 
\label{IVCMBN}
\end{figure}

Generally, the transport characteristics of a heterojunction depend upon the band lineups, temperature, majority/minority charge carriers, etc. The role of these parameters on the transport characteristics is  discussed in the following discussion. 

In our earlier work, we have reported the band offsets of Mg$_{0.21}$Zn$_{0.79}$O with respect to ZnO \cite{APL} as well as Ni$_{0.07}$Zn$_{0.93}$O with ZnO \cite{CAP}. According to the data from those reports, the conduction band and valence band offsets of Mg$_{0.21}$Zn$_{0.79}$O (Ni$_{0.07}$Zn$_{0.93}$O) with respect to ZnO is -0.1\,eV (0.2\,eV) and 0.3\,eV (0.35\,eV), respectively. The negative (positive) sign indicates that the corresponding band lies below (above) the band of ZnO. The positions of the valence band in Ni$_{0.07}$Zn$_{0.93}$O and Mg$_{0.21}$Zn$_{0.79}$O are also not same. Therefore one of the causes of nonlinear behavior may be attributed to the different band lineups. The band lineups may change due to the variation of temperature which can be determined from the knowledge of the Varshini parameters \cite{Varshini}. The Varshini parameters for the doped samples were not available. However, the magnitude of these parameters is very small ($\approx$ 10$^{-4}$\,eV/K) and we assume that the temperature dependent band gap variation is not responsible for the change in the transport characteristics.
%\begin{figure} %Mgcore
%\centering
%\includegraphics[width = 0.9\columnwidth]{I-VSch}
%\caption{Band diagram of Ni$_{0.07}$Zn$_{0.93}$O/Mg$_{0.21}$Zn$_{0.79}$O heterojunction at 10 K and 100 K.}% 
%\label{I-VSch}
%\end{figure}

Another possibility of non-ohmic behavior may be due to the change in the carrier concentration or majority type charge carriers of the junction. For this we have further carried out Hall measurements at 10\,K and 100\,K in single layer Ni$_{0.07}$Zn$_{0.93}$O film as shown in fig. 3. From fig. 3(a), it is evident that the grown samples exhibits positive Hall voltage at 10\,K while a negative Hall voltage is observed at 100\,K. The linearity of the Hall contacts is checked from the linear I-V behavior on Hall bar as shown in fig. 3(b) and 3(c) at 100\,K and 300\,K, respectively. Quite interestingly, we find a positive (10\,K) and negative (100\,K) Hall voltages in Ni$_{0.07}$Zn$_{0.93}$O film which show that the type of majority charge carriers changes with temperature. Due to high resistance of Mg$_{0.21}$Zn$_{0.79}$O sample, we were not able to perform its Hall measurement at low temperatures, however the room temperature measurements show that sample possesses an electron carrier concentration of the order of 10$^{19}$\,cm$^{-3}$ \cite{JAP}. The low temperature n-type behavior of MgZnO has also been confirmed by Pan \textit{et al.} in their temperature dependent Hall measurement studies \cite{Pan}.  
 
\begin{figure} %VBS
\centering
\includegraphics[width = 0.55\columnwidth]{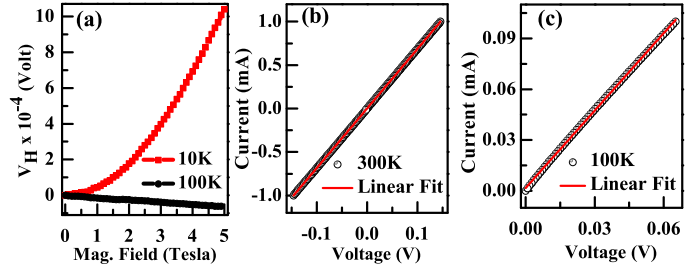}\hfill
\caption{(a) Behavior of Hall voltage of Ni$_{0.07}$Zn$_{0.93}$O sample with magnetic field, at 10\,K and 100\,K. (b) and (c) shows the linearity of the contacts on the Hall structure at 100\,K and 300\,K, respectively.}% 
\label{HV1}%
\end{figure} 

The origin of p-type conductivity in Ni$_{0.07}$Zn$_{0.93}$O at 10\,K and its transition to n-type at 100\,K can be understood by assuming that Ni doping induces vacancies in the valence band by trapping electrons at 10\,K. The trapping occurs because of the fact that Ni has higher electronegativity than Zn. The presence of Ni 3d states near the valence band has been confirmed from valence band spectroscopy measurements as shown in fig. (4). From the figure, it can be observed that feature appearing at 2\,eV corresponds to Ni 3d states in Ni$_{0.07}$Zn$_{0.93}$O thin film.
At 10\,K the electrons do not have sufficient energy to jump to the states available near conduction band but are trapped at Ni site. This leaves the vacancy in the valence band. As the temperature increases the electron concentration increases and due to high thermal energy are excited to higher energy states near conduction band, finally resulting in n-type conduction.
\begin{figure} %Mgcore
\centering
\includegraphics[width = 0.55\columnwidth]{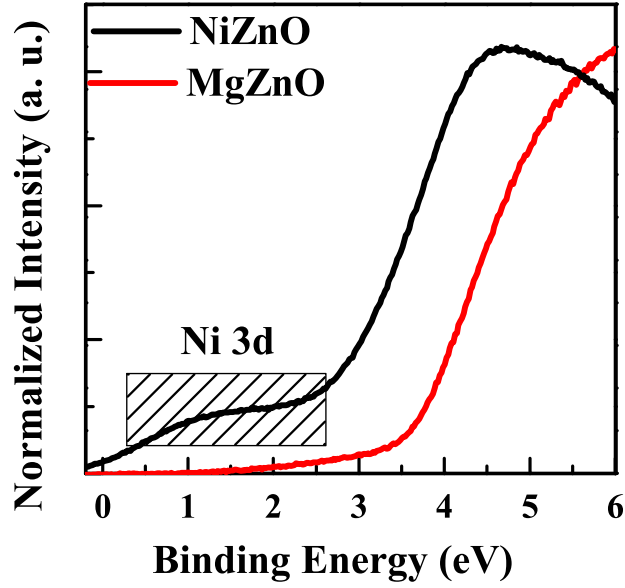}
\caption{Valence band spectra of Ni$_{0.07}$Zn$_{0.93}$O and Mg$_{0.21}$Zn$_{0.79}$O films.}% 
\label{vbs}
\end{figure}

 The possibility of creation of large electron concentration at higher temperature in Ni$_{0.07}$Zn$_{0.93}$O film can be explained on the basis of the following reported observation; The resonant photoemission spectroscopy studies by Yin et al. \cite{Yin} show that in Ni$_{0.14}$Zn$_{0.86}$O, Ni-3d states lie near the valence band of ZnO. This feature has also been observed in our recent study of the valence band spectra of NiZnO (Ni=5\%) \cite{TSF}. The Ni-3d states are split into lower energy doublet (e$_{g}$) and higher energy triplet (t$_{2g}$) states due to the crystal field splitting \cite{Shubhra}. The t$_{2g}$ states are further split into the low energy bonding states and high energy antibonding states. The antibonding states have higher energies and contain itinerant electrons. The energy of the antibonding states lies close to the conduction band of ZnO and hence with the increase in temperature, electrons in these states can jump easily into the conduction band because of thermal activation \cite{Shubhra}. However, amount of splitting leading to the proximity or overlapping of antibonding states with the conduction band depends upon the type and concentration of dopant ions. This feature explains n-type conduction at 100\,K in Ni$_{0.07}$Zn$_{0.93}$O. 
\begin{figure} %band
\centering
\includegraphics[width = 0.55\columnwidth]{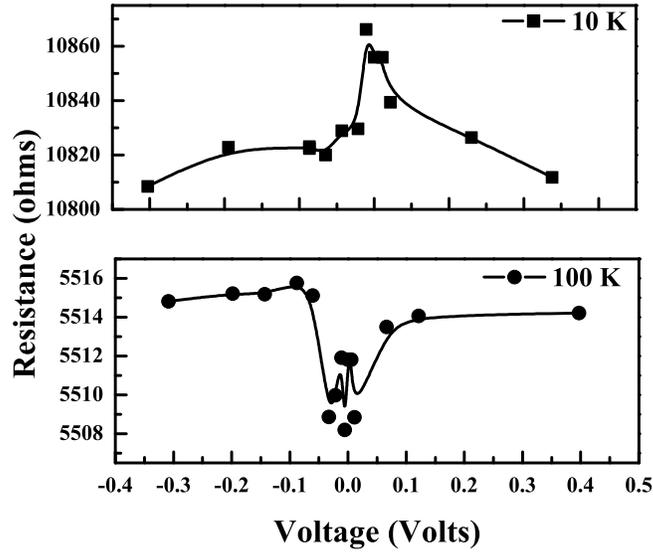}%
\caption{The Resistance-voltage relationship for Ni$_{0.07}$Zn$_{0.93}$O/Mg$_{0.21}$Zn$_{0.79}$O heterojunction at 10\,K and 100\,K.}% 
\label{RV}%
\end{figure}

Focusing our attention on ohmic behavior of I-V curve at 100\,K, we found that the calculated values of the carrier concentration in Mg$_{0.21}$Zn$_{0.79}$O (viz; 10$^{19}$\,cm$^{-3}$) and Ni$_{0.07}$Zn$_{0.93}$O (viz; 10$^{18}$\,cm$^{-3}$) do not vary significantly with temperature. Since the difference in carrier concentration at 100\,K in both Ni$_{0.07}$Zn$_{0.93}$O and Mg$_{0.21}$Zn$_{0.79}$O is very small. Since there is no sufficient concentration gradient, hence we do not observe non-ohmic behavior at the junction at these temperatures while at 10\,K the junction characteristics resemble with the p-n junction and cause non-ohmic behavior. We have explored the minute details of the junction characteristics by plotting resistance across the junction (R-V) near zero biasing regime in both forward and reverse bias conditions as shown in figure 5, where we find a strikingly opposite feature at 10\,K and 100\,K. At 100\,K the resistance is lower due to increased electron concentration. At 10\,K, since the junction acts as a p-n junction, so the formation of depletion region cannot be ignored at this temperature. The presence of this depletion region and low thermal energy results in the higher resistance of the grown heterojunction.
%At 100K our grown heterojunction is an n-n junction in which the conduction band edge of Ni$_{0.07}$Zn$_{0.93}$O is above that of Mg$_{0.21}$Zn$_{0.79}$O by an offset of the order of 0.3eV. Because of this, the charge carriers will experience a potential barrier during transport from Mg$_{0.21}$Zn$_{0.79}$O to Ni$_{0.07}$Zn$_{0.93}$ giving rise to a larger value of resistance. The charge carriers do not experience such barrier while moving from Ni$_{0.07}$Zn$_{0.93}$O to Mg$_{0.21}$Zn$_{0.79}$O. Consequently, the value of resistance is found to be higher in reverse bias while it is lower in the froward bias condition. Since the potential barrier height is only 0.3eV therefore this feature can be observed only near zero biasing regime. At 100K the majority charge carriers are electrons while at 10K majority charge carriers are holes which gives rise to opposite feature at 10K as shown (fig. 4). 

In conclusion, we find that Ni$_{0.07}$Zn$_{0.93}$O/Mg$_{0.21}$Zn$_{0.79}$O heterojunction acts as a p-n junction at low temperature, viz; 10\,K while at higher temperature (100\,K), it shows n-n junction behavior. The reason for the same has been ascribed to different band lineups as well as thermal activation of electrons to conduction band at higher temperature. 

Authors are grateful to Dr. M. Gupta, Dr. R. Rawat and Dr. A. Lakhani from UGC-DAE CSR Indore, for extending the XRD, transport and Hall measurement facilities, respectively. Financial support from UGC-DAE CSR, Indore and MPCST, Bhopal are highly acknowledged.


\begin{thebibliography}{99}
\bibitem{Ozgur}%1
U. Ozgur, Y. I. Alivov, C. Liu, A. Teke, M. A. Reshchikov, S. Dogan, V. Avrutin, S. J. Cho and H.A. Morkoc, J. Appl. Phys. \textbf{98}, (2005)  041301.
\bibitem{Look}
D. C. Look, Mat. Sci. Eng. B.-Adv. \textbf{80}, 2001 383–387.
\bibitem{Janoti}
 A. Janotti and C. G. Van de Walle, Appl. Phys. Lett. \textbf{87}, (2005) 122102.
\bibitem{Yoon}
J. -G. Yoon, S. W. Cho, W. S. Choi, D. Y. Kim, H. Chang, C. O. Kim, J. Lee, H. Jeon, S-H Choi and T. W. Noh, J. Phys. D: Appl. Phys. \textbf{44}, (2011) 415402.
\bibitem{A}%1
S. W. Tsang, M. W. Denhoff, Y. Tao and Z. H. Lu, Phys. Rev. B  \textbf{78}, (2008) 081304(R).
\bibitem{B}%2
S. H. Jeong, S. H. Song, K. Nagaich, S. A. Campbell and E. S. Aydil, Thin Solid Films \textbf{519}, (2011) 6619.
\bibitem{Edahiro}
T. Edahiro, N. Fujimura and T. Ito, J. Appl. Phys. \textbf{93}, (2003) 7673. 
\bibitem{C}%3
N. M. Kiasari, S. Soltanian, B. Gholamkhass and P. Servati, Sensors and Actuators A:Physical \textbf{182}, (2012) 105.
\bibitem{Saiket}%4
S. Chattopadhyay, P. Sen, J. T. Andrews and P. K. Sen, J. Appl. Phys. \textbf{111}, (2012) 034318.
%\bibitem{GP}%5a
%G. P. Agrawal, ``Fiber-Optic Communication Systems: 3rd Edition'', (John Wiley and Sons, USA, 2002), p. 107. 
\bibitem{JAP}%10
A. Agrawal, T. A. Dar, P. Sen and D. M. Phase, J. Appl. Phys. \textbf{115}, (2014) 143706.
\bibitem{Shubhra}
S. Singh and M. S. R. Rao, Phys. Rev. B \textbf{80}, (2009) 045220.
\bibitem{CAP}%12
T. A. Dar, A. Agrawal, P. Misra, L. M. Kukreja, P. K. Sen and P. Sen, Curr. Appl. Phys. \textbf{14}, (2014) 175.
\bibitem{APL}%13
A. Agrawal, T. A. Dar, D. M. Phase and P. Sen,  Appl. Phys. Lett. \textbf{105}, (2014) 081606. 
%\bibitem{Manish}%15
%M.K. Bafna, P. Sen and P.K. Sen, J. Appl. Phys., \textbf{100} (2006) 103522. 
\bibitem{Varshini}%16
Y. P. Varshni, Physica \textbf{34}, (2002) 154.
\bibitem{Pan}
X. H. Pan, W. Guo, Z. Z. Ye, B. Liu, Y. Che, W. Tian, D. G. Schlom and X. Q. Pan, Appl. Phys. Lett. \textbf{95}, (2009) 152105.
\bibitem{Yin}
Z. Yin, N. Chen, F. Yang, S. Song, C. Chai, J. Zhong, H. Qian and K. Ibrahim, Solid State Commun. \textbf{135}, (2005) 433.
\bibitem{TSF}
T. A. Dar, A. Agrawal, R. J. Choudhary and P. Sen, Thin Solid Films \textbf{589}, (2015) 821.
\bibitem{Monch}%17
W. Monch, Semiconductor Surfaces and Interfaces, (Springer-Verlag, Berlin Heidelberg, 1993), p. 79.%\bibitem{kittel}%18
%C. Kittel, Introduction to Solid State Physics: 8th Edition, (John Wiley and Sons, USA, 2005),p. 509.
\end{thebibliography}
\end{document}